\documentstyle[aps,multicol,prl,epsfig]{revtex}
\draft

\newcommand {\e}   {{\rm e}}
\newcommand {\eps} {\varepsilon}

\begin{document}

\draft
\title{Raft Instability of Biopolymer Gels\footnote{
Published in {\em Phys. Rev. Lett.} {\bf 87}, 158101 (2001)}}
\author{Itamar Borukhov$^{1,2}$ and Robijn F. Bruinsma$^{2,3}$}
\address{$^1$ Department of Chemistry and Biochemistry,
University of California at Los Angeles, Los Angeles, CA 90095-1569, USA}
\address{$^2$ Department of Physics,
University of California at Los Angeles, Los Angeles, CA 90095-1569, USA}
\address{$^3$ Instituut-Lorentz for Theoretical Physics,
Universiteit Leiden Postbus 9506, 2300 RA Leiden, The Netherlands}

\date{\today}

\maketitle

\begin{abstract}
Following recent X-ray diffraction experiments by Wong, Li, and Safinya on 
biopolymer gels, we apply Onsager excluded volume theory to a 
nematic mixture of rigid rods and strong ``$\pi/2$'' cross-linkers 
obtaining a long-ranged, highly anisotropic depletion attraction between 
the linkers. This attraction leads to breakdown of the percolation 
theory for this class of gels, to breakdown of Onsager's second-order 
virial method, and to formation of heterogeneities in the form of 
raft-like ribbons. 
\end{abstract}


\begin{multicols}{2}
\narrowtext

The demonstration by Lars Onsager in 1949 \cite{Onsager} that solutions of 
long thin rods undergo a first-order phase-transition from an isotropic to a 
birefringent nematic phase has remained a landmark achievement of theoretical 
statistical physics. He showed that at the isotropic/nematic transition point 
the rod volume fraction $\phi$ is surprisingly low, of the order of the 
aspect ratio $D/L$ of the rods, $D$ being the rod diameter and $L$ the rod 
length. 
Stiff biopolymers provide interesting applications of Onsager theory. 
In particular, the rod-like filamentous protein actin, which carries many 
biophysical functions \cite{ActinReview}, exhibits an isotropic-to-nematic 
transition \cite{ActinLC,Kas} at a critical volume fraction in approximate 
agreement with the Onsager criterion \cite{LC_lp}. Inside cells, micron-size 
actin filaments are part of a gel structure, the cytoskeleton, with the 
filaments cross-linked by reversibly bound proteins like $\alpha$-actinin. 
The unusual elastic properties of {\em in-vitro} actin gels 
\cite{Kas,Hinner} have recently been the focus of an intense 
theoretical effort \cite{ActinGelElasticity}. 

  {\em In-vitro} studies of the sol-gel transition of fixed-length 
actin filaments are currently interpreted in terms of percolation theory 
\cite{ActinPerculation}, in which linkers are assigned, at random, to 
rods providing connections to neighboring rods with 
{\em no preferred crossing angle}. The sol-gel point, 
where the real and imaginary parts of the elastic moduli are
comparable in magnitude, is identified as the 
percolation threshold, i.e., the linker concentration for 
which a connected ``path'' of linked rods stretches across the sample 
for the first time. 
An important validity condition is that the concentration of 
``native'' rod-rod contacts - i.e., the mean number density of rod-rod 
contacts {\em before} linkers are added - exceeds the percolation threshold, 
otherwise the reduction in configurational freedom of the rods imposed by 
the linkers would lead to phase-separation. The number of native contacts per 
rod in a dilute isotropic solution of rods is of the order of $(L/D)\phi$ 
so a rod-linker solution with $\phi$ of order $D/L$ is expected to exhibit 
a sol-gel transition that is reasonably well described by percolation theory. 
The same argument indicates that the structure factor $S({\bf q})$ 
should resemble that of a pure Onsager nematic.

\begin{figure}[b]
\vfill
\centerline{\epsfxsize=3.4in \epsfbox{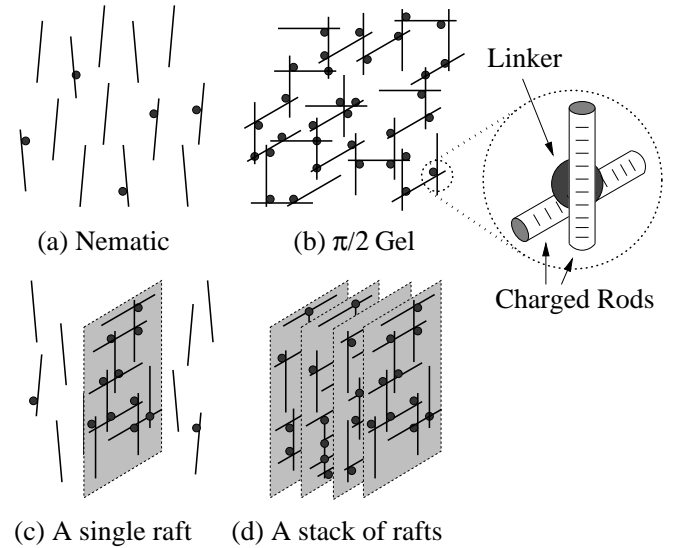}}
\vspace{\baselineskip}
\caption{\protect\footnotesize Mixtures of rigid rods in the nematic phase 
and ``$\pi/2$'' linkers display a complex range of structures. 
(a) At very low linker concentrations and binding energy, 
excluded volume effects 
prevent cross-linking. 
(b) Second-virial theory predicts ``$\pi/2$'' tetratic or cubatic gels at 
high linker concentrations.
(c) At low linker concentrations a long-range depletion attraction leads 
to heterogeneous structures in the form of raft-like ribbons along the optical 
axis. 
(d) Rafts can stack on top of each other as a result of inter-raft attraction 
leading to phase separation of the linker-rich stacks from the 
nematic phase.
}
\label{fig:networks}
\end{figure}

   Recent high-precision low-angle synchrotron X-ray studies \cite{Wong} 
of the microscopic properties of mixtures of actin filaments and positive 
divalent counterions report results that are in conflict with these 
expectations. When divalent counterions are added to a 
nematic phase of actin filaments, a birefringent gel phase is observed at 
linker concentrations that depends sensitively on the (average) rod length 
$L$, whereas percolation theory would predict that, beyond a certain length, 
the critical linker concentration is determined only by the rod concentration.
In addition, {\em a harmonic sequence} is found in the structure 
factor $S({\bf q})$ of the gel phase in the long-wavelength range 
$1/L\alt q\alt 1/D$ along directions approximately perpendicular to the 
optical axis. This is incompatible with the Onsager nematic 
\cite{vanderSchootComm}, but it {\em is} typical of layered structures, 
such as smectic liquid crystals. 
The X-ray experiments on actin gels ­ as well as similar results obtained 
on other biopolymers such as DNA \cite{SafinyaComm} ­ clearly conflict 
with a simple ``Onsager-Percolation'' type description. 

   It is the claim of this paper, that the unusual features of biopolymer 
gels can be physically interpreted if we allow for the fact that
linked biopolymers have preferential crossing angles.
First, actin linker proteins are known to impose different crossing angles, 
depending on their molecular structure.
Second, biopolymers like actin or 
DNA carry a high linear charge density, which allows them to be water soluble. 
The electrostatic repulsion between two uniformly charged rods 
linked at a single point depends on the relative angle $\gamma$ as 
$1/|\sin\gamma|$ \cite{Brenner}, so the optimal 
crossing angle between two linked biopolymers is, in general, expected to be 
large. Based on this consideration, we studied 
- as a simple model for biopolymer gel formation - 
a dilute mixture of rigid rods with a concentration $\rho_p$ of the order of
 $1/DL^2$, in the presence of a concentration $\rho_l$ of freely sliding, 
reversible linkers connecting pairs of rods at $\pi/2$ crossing angles 
(binding energy $\eps_0$). When $\rho_l=0$, second-virial theory predicts a 
first-order phase transition from an isotropic to a nematic phase. 
It is easy to demonstrate \cite{Bruinsma} that at non-zero linker 
concentrations, second-virial theory predicts a first-order 
phase transition (for $\rho_l\propto\rho_p$) from the nematic phase 
to a birefringent ``$\pi/2$'' gel phase with tetragonal symmetry 
\cite{Tetragonal} (see Fig.~\ref{fig:networks}b). 

\begin{figure}
\centerline{\epsfxsize=3.0in \epsfbox{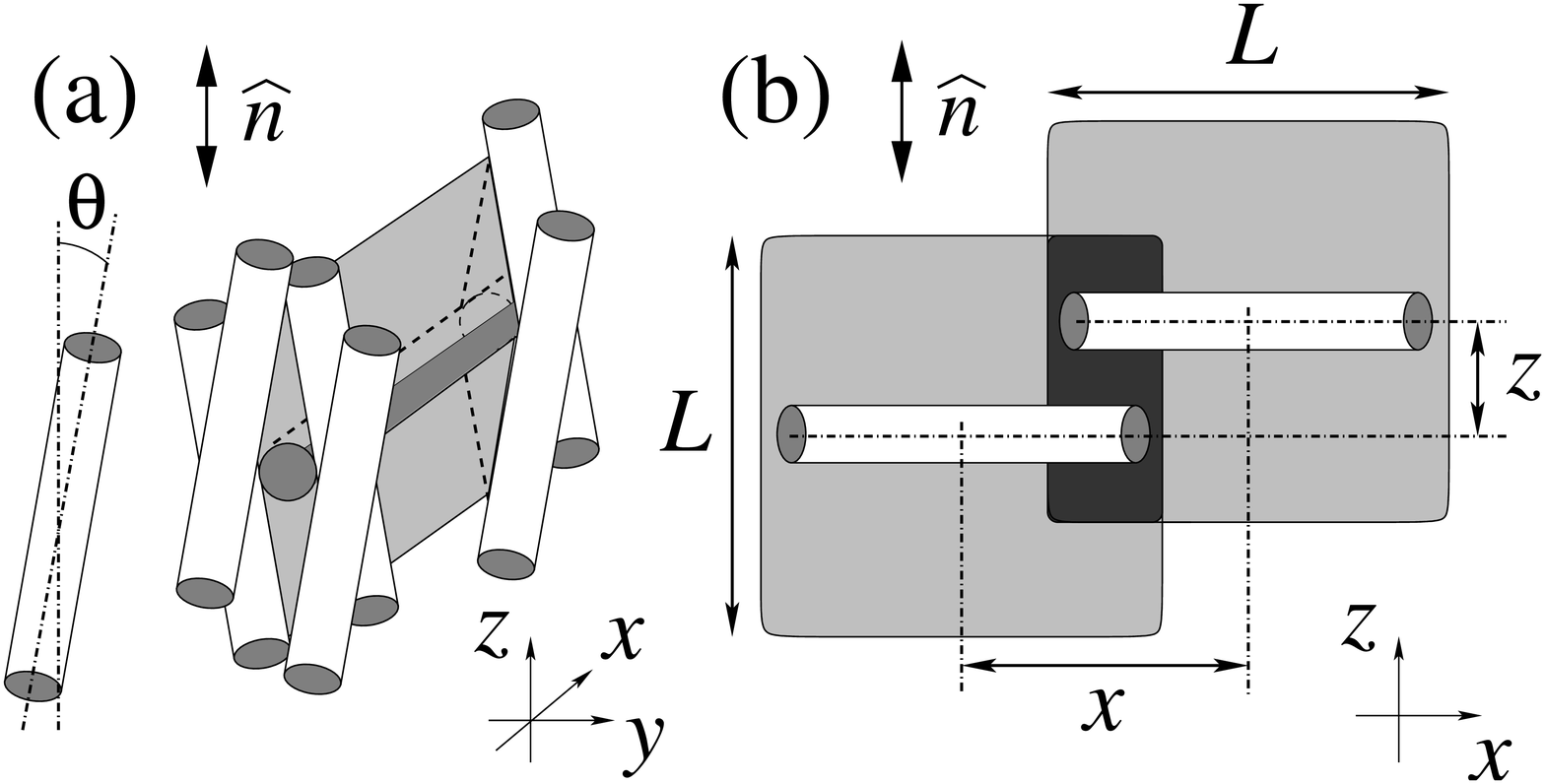}}
\vspace{\baselineskip}
\caption{\protect\footnotesize 
Schematic view of 
(a) the depletion volume of a transverse rod in an Onsager nematic 
and (b) the overlap of two depletion volumes surrounding two 
transverse rods for ${\bf R}=(x,0,z)$ and  $\gamma_{tt}=0$.
}
\label{fig:shadow}
\end{figure}

   To examine gel formation beyond second-virial theory, 
we follow what happens when linkers are added, one by one, 
to the rod solution. 
A single linker connecting two rods produces a cross-like structure, 
with the two rods free to slide with respect to each other. 
It is easy to show that the free energy has a minimum when one arm of the 
cross extends along the optical axis (the $z$-direction) 
with the other arm taken to be the transverse $x$-direction \cite{45degrees}. 
As shown in Fig.~\ref{fig:shadow}a, the transverse rod creates an 
anisotropic depletion volume $\delta V_1$ for the centers of mass (CM's) 
of the free rods. This depletion volume depends on the angle $\gamma_{ft}$ 
between the axis of a free rod and the axis of the transverse rod by 
the Onsager result 
$\delta V_1(\gamma_{ft})\simeq 2DL^2|\sin\gamma_{ft}|$. 
The entropic self-energy of the linked cross is computed as the osmotic work
$\delta W_1\simeq\Pi_{\rm osm}\langle \delta V_1(\gamma_{ft})\rangle_\theta$
required to remove the CM's of the free rods from the depletion volume, 
with $\Pi_{\rm osm}$ the osmotic pressure of the nematic 
and with $\langle~\rangle_\theta$ denoting an orientational average. 
The effect of the arm of the cross along the optical axis is smaller by a 
factor $D/L$ and will be neglected. 
Using Onsager theory for the orientational distribution of the free rods, 
we find:
\begin{equation}
  {\delta W_1\over k_BT} \simeq \left\{ 
\begin{array}{ccl}
 2c_I(1+c_I) && ({\rm Isotropic,}~ c_I\alt 3.3) \\ 
 (24/\pi)c_N(1-{\Delta\theta^2/4}) && ({\rm Nematic,}~ c_N\agt 4.5) \\ 
\end{array}
\right.
\label{eq:dW1}
\end{equation}
with $\Delta\theta^2$ being the (small) variance of the polar angle with 
respect to the optical axis, 
and $c=(\pi/4)DL^2\rho_p$ the Onsager dimensionless rod concentration. 
At the isotropic/nematic transition point, the dimensionless free energy cost
${\delta W_1/k_BT}\simeq 33.2$
assumes a large value independent of the dimensions of the rod, 
or the rod concentration. 

   Now add a second linker with a second rod making a $\pi/2$ angle 
with the optical axis (see example in Fig.~\ref{fig:shadow}b). 
Let the relative distance between the CM's of the two transverse rods 
be ${\bf R}$, let the angle of the second rod with the $x$ direction 
(i.e., the relative angle) be $\gamma_{tt}$, 
and let $\delta V_2({\bf R},\gamma_{tt})$ be the excluded volume 
that is {\em shared} between the two rods. 
The entropic free energy cost of the two-rod system will then be
\begin{equation}
   \delta W_2({\bf R},\gamma_{tt}) \simeq 
2\delta W_1-\Pi_{\rm osm}\langle \delta V_2({\bf R},\gamma_{tt})\rangle_\theta
\label{eq:dW2gen}
\end{equation}
In the limit $R\equiv|{\bf R}|\to 0$ and $\gamma_{tt}=0$, 
the two rods coincide so $\delta W_2(0,0)\simeq\delta W_1$. 
On the other hand, for $R\ge L$, there is no shared excluded volume so 
$\delta W_2({\bf R},\gamma_{tt})\simeq2\delta W_1$. 
It follows that the two linkers attract each other with a potential energy 
that has a typical range of order $L$ 
and a typical binding energy of order $\delta W_1$. 
This effect is related to the well-known {\em depletion attraction} 
\cite{depletion}, the attraction between large objects in a surrounding 
medium of small objects driven by the increase in configurational entropy 
as the larger objects coagulate. 
The difference is that in the present case an effective attraction between 
small objects (linkers) is generated by an increase in configurational 
entropy of large objects (rods).
Using once again the Onsager orientational 
distribution, we find that, according to Eq.~\ref{eq:dW2gen}, 
the linker-linker depletion attraction drops off inversely proportional to 
the distance between the rods: 
\begin{equation}
   {\delta W_2(z,\gamma_{tt})-2\delta W_1 \over \delta W_1} \simeq
	-a_1 {D\over\Delta\theta|z|} 
        \min\left({D\over L|\sin\gamma_{tt}|},1\right)
\label{eq:dW2a}
\end{equation}
where $z$ is the separation between the rods along the optical axis and 
 $a_1\simeq 6/\sqrt{\pi}$. 
The interaction is highly anisotropic: for small $|z|$, 
there is a strong torque between transverse rods favoring parallel alignment 
(i.e., $\gamma_{tt}=0$). Moreover, the mechanism operates 
only as long as the CM of the second rod lies below or above the first rod 
within the two narrow wedges bordered by $|y|\simeq\Delta\theta|z|$ 
centered on the first rod. 

   Equation~\ref{eq:dW2a} is valid for $D\ll \Delta\theta|z|\ll L$. 
For small separations $\Delta\theta|z|\ll D$ the potential has 
a linear dependence on $z$:
\begin{eqnarray}
   {\delta W_2(z,0)-2\delta W_1 \over \delta W_1} \simeq && \nonumber\\
	-\left( 1-a_2 {\Delta\theta|z|\over D} \right) &&
        \min\left({D\over L|\sin\gamma_{tt}|},1\right)
\label{eq:dW2b}
\end{eqnarray}
with $a_2\simeq 1/2\sqrt{\pi}$. 
This result can be understood by noting that for perfectly aligned rods 
the shared depletion volume of two transverse rods would have a 
linear dependence on separation \cite{vanderSchoot}. 

\begin{figure}
\centerline{\epsfxsize=3.0in \epsfbox{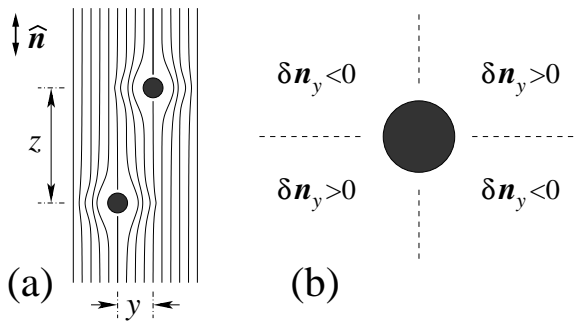}}
\vspace{\baselineskip}
\caption{\protect\footnotesize (a) Schematic view of the long-range 
depletion interaction between two transverse rods. 
The continuous lines indicate the director field. 
(b) The quadrupole symmetry of the distortion zone around a transverse rod. 
}
\label{fig:continuum}
\end{figure}

\begin{figure}
\centerline{\epsfxsize=3.0in \epsfbox{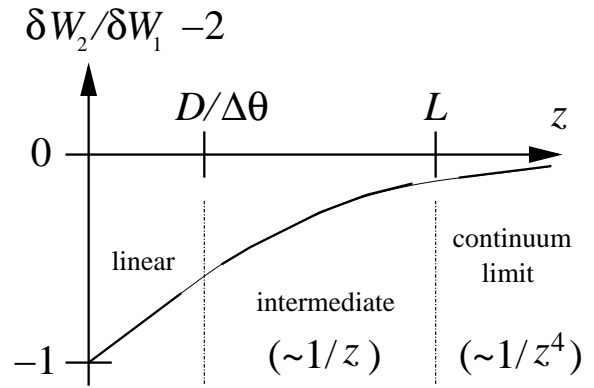}}
\vspace{\baselineskip}
\caption{\protect\footnotesize 
The depletion attraction between the two rods for $y=0$. 
}
\label{fig:dW}
\end{figure}

   In the opposite limit of rod separations large compared to the 
rod length $L$, the continuum description of a nematic \cite{Poulin} 
can be used in terms of a locally varying director field 
$\hat{\bf n}({\bf r})$ (see Fig.~\ref{fig:continuum}a). 
The perturbation of $\hat{\bf n}({\bf r})$ far from the fixed rod can be 
written as $\hat{\bf n}({\bf r})\simeq (0,\delta n_y({\bf r}),1)$, 
where $\delta n_y\ll 1$. 
The leading term in the elastic free energy in this geometry 
is the {\em splay} term: 
\begin{equation}
   F_{\rm splay} = {K_{11}\over 2}\int{\rm d}^3r (\nabla\delta n_y)^2
\label{eq:Fsplay:gen}
\end{equation}
where $K_{11}$ is the splay elastic constant and 
 is equal to $K_{11}\simeq 0.06DL^4\rho_p^2k_BT$ 
for rigid rods \cite{K11}.
Treating $\delta n_y({\bf r})$ as a variational function and minimizing 
Eq.~\ref{eq:Fsplay:gen} leads to the Laplace equation 
 $\nabla^2\delta n_y=0$. 
As demonstrated in Fig.~\ref{fig:continuum}b, the appropriate solution of the 
Laplace equation must have the symmetry of a two-dimensional {\em quadrupole} 
so $\delta n_y(y,z)=Q yz/(y^2+z^2)^2$. 
The value of the quadrupole moment $Q$ is determined by the 
condition that the typical deviation $Q/L^2$ of $\delta n_y({\bf r})$ for
 $y^2+z^2\simeq L^2$ should match the angular deviation $D/L$ of 
the free rods as obtained from excluded volume arguments 
(see Fig.~\ref{fig:shadow}a). This condition leads to 
 $Q\propto DL$ times a function of $\Delta\theta$. 
From the electrostatic interaction energy between quadrupoles in 
two dimensions, $Q^2\cos(4\varphi)/r^4$ (in polar coordinates $r,\varphi$), 
it follows that the interaction energy per unit length between two parallel 
(infinite) rods must have the form:
\begin{equation}
{F_{\rm splay}\over L k_BT} \simeq 
   g(\Delta\theta)\cos(4\varphi) c^2 {DL^2\over r^4}
   ~~~r\gg L
\label{eq:Fsplay}
\end{equation}
where the function $g(\Delta\theta)$ includes the dependence on the 
nematic order parameter. The deformation energy $F_{\rm splay}$ 
should be matched with the right hand side of Eq.~\ref{eq:dW2gen} at distances 
 $r\simeq L$ yielding $g\propto 1/c$. 
It follows from Eq.~\ref{eq:Fsplay} that the anisotropy of the linker-linker 
interaction is significantly less pronounced in the regime $r\gg L$. 
Moreover, the interaction is {\em repulsive} when the separation vector 
between the CM's of the rods makes an angle near $\pi/4$ 
with the optical axis. 
The various regimes are shown schematically in Fig.~\ref{fig:dW}. 

   We now turn to the implication of these results. 
According to Eq.~\ref{eq:dW1}, the characteristic energy scale $\delta W_1$ 
of the long-range depletion attraction between $\pi/2$ linkers is 
large compared to the thermal energy $k_BT$. 
This implies that percolation theory is not applicable since linkers 
will cluster even at very low concentrations. 
Next, second-virial theory is not reliable either since it gives no 
indication of this form of phase separation. Indeed, the cross-links 
resemble flat square discs of dimensions $L\times L$ and second-virial 
expansions are known not be accurate for mixtures of rods and discs
\cite{vanderKooij}. 
The large value of $\delta W_1/k_BT$ would have another important consequence:
unless the linker binding energy is sufficiently high, 
{\em crosses simply would not be able to form}. 
Assuming chemical equilibrium between linkers in 
solution, linkers adsorbed on isolated chains, and linkers forming 
a cross-structure, we find that the fraction $f$ of linkers able to form 
a cross is of order:
\begin{equation}
   f \simeq {1\over 1+(\pi/2c)\e^{(\eps_0-\delta W_1)/k_BT}}\
\label{eq:fcrosslinks}
\end{equation}
This means that cross-links disappear when $\delta W_1$ is large compared 
to $\eps_0$, i.e., when rods are so long that $DL^2\gg \eps_0/k_BT\rho_p$. 
This is consistent with the experimental observation that the critical linker 
concentration for actin gelation rapidly increases with rod length. 
For an actin solution with $\rho_p\simeq$0.1mM, 
a linker concentration $\rho_l\simeq$1mM and $\eps_0\simeq 15k_BT$, 
the rod length should not exceed a value of order 100nm. 

   What would be the structure of the clusters? 
According to Eqs.~\ref{eq:dW2a},~\ref{eq:dW2b}, 
the depletion attraction between 
two transverse rods is highly anisotropic at short distances and maximal 
if the two rods are parallel and located just below each other with a 
common plane whose normal is perpendicular to the optical axis. 
Adding additional linkers and transverse rods in the same way produces 
an elongated ribbon structure resembling a disordered {\em raft} 
(see Fig.~\ref{fig:networks}c). 
The sol-gel transition point should be characterized, not by percolation, 
but by the formation of elongated, ribbon-like rafts extending through 
the sample along the optical axis. 
Because two parallel plates immersed in a nematic attract each other 
\cite{Ajdari} - with a force per unit area that depends on the inter-layer 
spacing $h$ as $k_BT/h^3$ for large $h$ - 
depletion attraction should also operate between {\em different} rafts, 
so the ribbons could be several layers thick (Fig.~\ref{fig:networks}d). 
A multilayer ribbon-raft indeed would have along the transverse 
direction a structure factor $S({\bf q})$ similar to that of 
of a smectic, although without long-range order. 
Along the optical axis, there would have to be a peak in $S(q_z)$ 
when $2\pi/q_z$ equals the spacing between transverse rods inside the raft, 
which is indeed present \cite{SafinyaComm}.
	 
   Our method only applies to a small number of linkers. 
At finite linker concentration, rafts and ribbons are expected to act 
as long-lived kinetic intermediates. The actual thermodynamic equilibrium 
state is expected to be either a $\pi/2$ gel or a biaxial nematic 
\cite{vanderSchootComm,SafinyaComm}, in phase-coexistence with
linker-poor solution.
Numerical simulation methods of the model should be able to address this issue.
Next, we did not allow for the fact that linkers may prefer the endpoints 
of rods nor deviations away from $\pi/2$ for the preferential crossing angle. 
Finally, actin cytoskeletons {\em do} appear to resemble $\pi/2$ gels 
\cite{Janmey} with no evidence of raft formation. 
Actin filaments inside cells do not have a fixed length 
however: active polymerization and depolymerization processes are 
constantly taking place and {\em in-vitro} experiments of gel formation 
by such ``living'' actin filaments would be of great interest.


{\em Acknowledgments:} We would like to thank C. Safinya for communicating 
the results of the X-ray experiments on Actin gels as well as 
extensive discussions. We would like to thank P. van der Schoot and G.Wong 
for a critical reading of the manuscript and W. Gelbart, A. Liu, 
T. Lubensky and C. O'hern for useful discussions. 


\end{multicols}


\begin{references}
\bibitem{45degrees}
  A second minimum is encountered with two rods tilted by 45$^\circ$ 
with respect to the optical axis. However, the linker entropy favors 
the 0$^\circ$ configuration.
\end{references}
\end{document}